\documentclass[%
  twocolumn,%
  floatfix,%
  showpacs,
  twocolumn,%
  prb
]{revtex4}

\usepackage{amsmath,amssymb,amsxtra,bm,bbm}
\usepackage[dvips]{graphicx}
\usepackage{hyperref}
  \hypersetup{%
    breaklinks = {true},
    citecolor = {blue},
    colorlinks = {true},
    linkcolor = {red},
    pdfauthor = {\textcopyright\ Vitor M. Pereira},
    pdfcreator = {\LaTeX\ and \flqq hyperref\frqq},
    pdffitwindow = {true},
    pdfmenubar = {true},
    pdfpagelayout = {SinglePage},
    pdfstartview = {Fit},
    pdftoolbar = {true},
    plainpages = {false},
  }
\newcommand{\br}{\textbf{r}}
\newcommand{\bq}{\textbf{q}}
\newcommand{\bk}{\textbf{k}}

\begin{document}

\title{Polarization Charge Distribution in Gapped Graphene}

\author{Valeri N. Kotov}
\affiliation{Department of Physics, Boston University, 
590 Commonwealth Avenue, Boston, Massachusetts 02215}

\author{Vitor M. Pereira}
\affiliation{Department of Physics, Boston University, 
590 Commonwealth Avenue, Boston, Massachusetts 02215}

\author{Bruno Uchoa}
\affiliation{Department of Physics, Boston University, 
590 Commonwealth Avenue, Boston, Massachusetts 02215}


\begin{abstract}
We study the distribution of vacuum polarization charge induced
 by a Coulomb impurity in massive graphene.
By analytically computing the  polarization function, we show that the 
charge density is distributed in space in a non-trivial fashion, and on a
characteristic length-scale set by the effective Compton wavelength. 
The density crosses over from a logarithmic behavior below this scale, to a power
law variation above it. 
Our results in the continuum limit are confirmed by explicit
diagonalization of the corresponding tight-binding model on a finite-size lattice. 
Electron-electron interaction effects are also discussed.
\end{abstract}

\pacs{81.05.Uw, 73.43.Cd}


\maketitle

 
%
\section{Introduction and formulation of the problem}
Over the course of the past year the behavior of graphene in the presence of a
strong external Coulomb field was analyzed in considerable detail. 
\cite{vitor1,levitov,sachdev,ivan,fogler,vitor2} This problem is important,
notably, for our understanding of electronic transport in the presence
of charged impurities.\cite{antonio} In addition, since
the effective coupling constant in graphene, $\alpha= e^2/(\hbar v)$,
can be rather large (in vacuum $\alpha = 2.2$), the exploration of features
uniquely associated with the strong-field regime $Z\alpha \sim 1$ (where $Z$
stands for the strength of the external Coulomb field: $V(r) = Z e /r$) becomes
a realistic prospect. In this context, it was found that above a critical
value $(Z\alpha)_c = 1/2$ characteristic resonances appear in the energy
spectrum \cite{vitor1,levitov} and the vacuum polarization density varies as
$\rho({\br}) \sim 1/r^{2}$. In the subcritical regime ($Z\alpha<1/2$), on the
other hand, the polarization charge is concentrated around the
Coulomb center in such a way that one obtains $\rho({\br}) \propto
\delta({\br})$, within the continuum approximation for the electron dynamics.
The physics around the critical point $(Z\alpha)_c$ is a peculiar massless
realization of the more ``conventional" vacuum charging behavior in massive
quantum electrodynamics (QED),\cite{Greiner} which also seems possible in
graphene under certain conditions.\cite{vitor2}

The drive to explore unconventional behavior at strong coupling
($Z\alpha \sim 1$) has been fueled, so far, by theoretical progress only.
Experiments in which K$^+$ ions are deposited in a controlled way onto graphene
show the behavior one expects from the theory of scattering of Dirac fermions by
a Coulomb field,\cite{fuhrer,novikov} but only in the undercritical regime, 
perhaps as expected for such low value of $Z$. It is also clear that under
current experimental conditions $\alpha < 1$ due to dielectric screening by
the substrate, and additional screening is provided by the presence of 
water layers around the samples.\cite{schedin} This conspires to 
significantly increase the effective dielectric constant with a
concomitant decrease of $\alpha$, thus making the subcritical regime
$Z\alpha < (Z\alpha)_c$ the relevant one for present-day experiments, and also
accounting  for the feeble signatures of interaction effects.
It is therefore natural to ask how the characteristic 
features of the undercritical regime  manifest themselves in the vacuum polarization and
screening properties. In the strict massless limit the polarization charge density is
concentrated at the potential source, $\rho({\br}) \propto  Z\alpha\delta({\bf r})$, and
non-trivial spatial variation can only occur  due to additional interaction
effects.\cite{sachdev} These are expected to be small, and we
will see that, perturbatively, they read $\rho(\br) \sim Z\alpha^2/r^2$.

In the present work we explore another source of density variation caused by
the presence of a finite ``mass'' $m$ or, equivalently, an energy gap $\Delta =
2 m v^2$ in the electronic spectrum. There are at least three  sources of a gap
in graphene. Firstly, it has
been realized recently \cite{lanzara} that epitaxially grown graphene exhibits a
substantial gap ($\Delta = 0.26 ~\text{eV}$) due to the breaking of the
sublattice symmetry by the substrate.
Graphene suspended above a graphite substrate also has a small gap 
$\Delta \approx 10 ~\text{meV}$.\cite{andrei}
 Secondly, spin-orbit coupling leads to a
gap, albeit of much smaller magnitude: $\Delta_{so} \simeq
10^{-3}~\text{meV}$.\cite{so}
Finally, real mesoscopic samples can have an effective gap generated by
their finite size, which scales as $\Delta \sim 1/L$.
We find that the polarization density of massive Dirac fermions displays
 characteristic behavior, controlled by the effective ``Compton'' wavelength
$\lambda_C = \hbar/(mv)$. Our main results are summarized in Figs.~\ref{Fig2}
and \ref{Fig3}: for distances $r\lesssim\lambda_C$, the
density variation is logarithmic, crossing over to a power law tail at
$r\gtrsim\lambda_C$. It should be possible to explore this behavior with modern
experimental techniques for detection of local density variation.\cite{yacoby,stolyarova,andrei} 

\begin{figure}[tb]
  \centering
  \includegraphics*[width=0.33\textwidth]{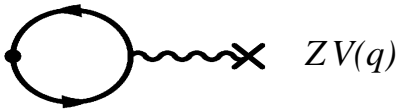}
  \caption{Graphical representation of Eq.~\eqref{ic}.}
  \label{Fig0}
\end{figure}

Our starting point is a 2D system of massive Dirac electrons in the presence of a
Coulomb impurity with effective charge $Ze$. The Hamiltonian is:
\begin{equation}
  \label{ham}
  \hat{H} =  - i\hbar\,v\,\bm{\sigma}\cdot \bm{\nabla}+ m v^2\sigma_{3} -
  \frac{Ze^2}{r} 
  \,,
\end{equation}
where $v$ is the Fermi velocity and $\sigma_i$ are the Pauli matrices. The induced
vacuum polarization charge is calculated in linear response according to 
\begin{equation}
  \label{ic}
  \rho(\bq) = ZeV(\bq)\Pi(\bq,0)
  \,,\quad V(\bq) = \frac{2\pi\alpha}{|\bq|},
\end{equation}
where $\alpha= e^2/(\hbar v)$ and $\Pi({\bq},0)$ is the static
polarization function. This equation is schematically represented by the diagram in  Fig.~\ref{Fig0}.
%
%

Unless specified otherwise, we  use units in which $v=\hbar=1$, the
electron charge is $-e$ ($e>0$) and, for convenience, we measure all charges 
($\rho,Q$) in units of $e$ (the sign is meaningful though). The chemical
potential, $\mu$, is assumed to be in the gap, $|\mu| < m$ (further discussion
appears later).

In the rest of the paper, we  first compute the polarization operator 
for massive graphene (Section II), which we use in Section III for the
 calculation of the induced vacuum polarization charge in a weak Coulomb field. 
 In Section IV we compare the obtained behavior  with results based on exact diagonalization studies
 on a finite-size lattice. Section V discusses the influence of weak electron-election
 interactions on the polarization charge, and Section VI contains our conclusions. 
%

%

%
\section{Polarization Function for Massive Dirac Particles}
The polarization function is computed in the standard way as
\begin{equation}
  \label{pol1}
  \Pi({\bf q},\omega) = -i N \sum_{\bf{k}} \int \frac{d\nu}{2\pi} {\mbox{Tr}} \{
  \hat{G}({\bf k},\nu)
  \hat{G}({\bf k + q},\nu+\omega) \},
\end{equation}
where the trace is over the Pauli matrices and $N=4$ accounts for the valley
and spin degeneracies. The Green's function at finite mass is given by
\begin{equation}
  \label{GF}
  \hat{G}(\bk,\nu) = \frac{\nu + {\bm \sigma}\cdot\bk +
  m\, \sigma_3}{\nu^2-\bk^2 - m^2  + i \eta} \ .
\end{equation}
Using a more symmetric 3-vector notation, $({\bf q},q_0) = ({\bf q},\omega)$,
$({\bf k},k_0) = ({\bf k},\nu)$, $k^{2} = {\bf k}^2 - k_0^2$, etc., we obtain
\begin{equation}
\label{polint}
\Pi({\bf q},q_0) = -8i \int \frac{d^3k}{(2\pi)^{3}} 
\frac{k_0(k_0 + q_0)+{\bf k}\cdot({\bf k + q}) + m^2}{(k^2+m^2)\bigl[(k+q)^2+m^2\bigr]}.
\end{equation}

Technically, an exact analytical evaluation of  \eqref{polint} becomes possible
if one treats the frequency and momentum integrations on equal footing, as
would happen in a Lorentz invariant situation, and was done for the massless
case.\cite{son} However in this way one encounters a linearly divergent piece,
 since it is clear that at large momenta Eq.~\eqref{polint} leads to
 $\int^{\Lambda} d^3k/k^2 \sim \Lambda$, where $\Lambda$ is the covariant ultraviolet cut-off.
 This procedure therefore requires ``regularization", i.e. subtraction of the infinite contribution.
The regularization procedure, however, yields the correct result because in a
non-relativistic situation, when the energy ($k_0$) integration is performed first in the 
interval $(-\infty,\infty)$, the resulting momentum integration is ultraviolet convergent
 (and cut-off independent).

\begin{figure}[tb]
  \centering
  \includegraphics*[width=0.47\textwidth]{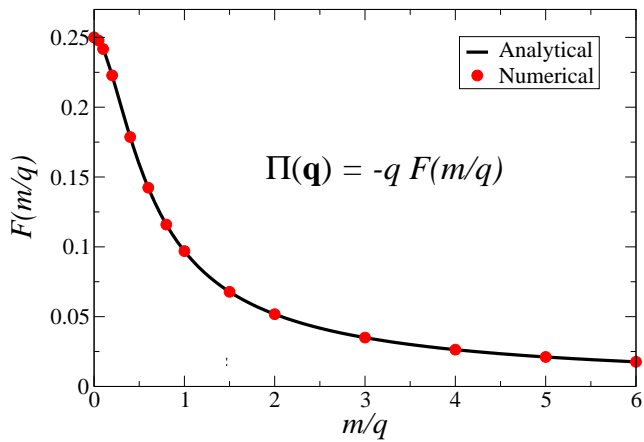}
  \caption{(Color online)
    Plot of the function $F(m/q)$, defined as: 
    $\Pi({\bf q},\omega=0) = -q F(m/q)$, where $q=|{\bf q}|$. 
  }
  \label{Fig1}
\end{figure}

Restoring the original notation,  we obtain the final exact expression for the polarization
\begin{equation}
  \label{pol3}
  \Pi(\bq,\omega)= -\frac{|{\bf q}|^2}{\pi} 
    \Biggl\{
      \frac{m}{q^2} + \frac{1}{2q}
        \biggl(1 -  \frac{4m^2}{q^2} \biggr) 
      \arctan{\Bigl(\frac{q}{2m}\Bigr)} 
  \Biggr\},
\end{equation}
where
\begin{equation}
 q=\sqrt{|{\bf q}|^2-\omega^2} \ .
\end{equation} 
Unlike QED, the polarization function of graphene is not covariant.
 We have confirmed this result by direct numerical integration of \eqref{pol1}.
A comparison of the two in the static case ($\omega=0$) is shown in Fig.~\ref{Fig1}, where the 
correspondence between the numerical calculation and Eq.~\eqref{pol3} is exact.

 At finite frequency, dynamical properties such as the longitudinal conductivity of gapped 
graphene can be derived directly from Eq.~\eqref{pol3}. For the real part of the conductivity
 $\sigma_m(\omega)$  one can use the standard formula,
  $\sigma_m(\omega) = -e^2 (\omega/{\bf q}^2){\mbox{ Im} }\Pi(\bq,\omega)$, \cite{EM} for
 $|{\bf q}|\rightarrow 0$. We readily extract
 the imaginary part of the dynamical polarization,
\begin{equation}
\label{impol}
{\mbox{ Im} }\Pi(\bq,\omega) \!=\! \frac{-|{\bf q}|^2}{4\tilde{q}}\left\{ 1+ 
\frac{4m^{2}}{\tilde{q}^2} \right \}\theta\!\left(\omega \! - \! \sqrt{|{\bf q}|^2 + 4m^{2}}\right),
\end{equation}
where we define
\begin{equation}
 \tilde{q}\equiv\sqrt{\omega^2- |{\bf q}|^2} \ .
\end{equation} 
 This leads to
\begin{equation}
\label{cond}
 \frac{\sigma_m(\omega)}{\sigma_0} = \left(1+ \frac{4m^{2}}{\omega^2} \right)\theta(\omega - 2m) \ , 
\end{equation}
where $\sigma_0$ is the conductivity of massless graphene, predicted to be $\sigma_0=e^2/(4\hbar)$.\cite{antonio}
 Eq.~\eqref{cond} implies that, at the edge $\omega=2m$, the conductivity  increases by a factor
 of two, $\sigma_m = 2 \sigma_0$, and decreases for higher frequencies, approaching
$\sigma_m =  \sigma_0$ ($\omega \gg m$). For a gap of $\Delta = 2m \approx 260 \ \mbox{meV}$,\cite{lanzara}
 we estimate that the characteristic frequencies where the enhancement should
 be observable are $\omega \sim \Delta \sim 2\times 10^{3} \ \mbox{cm}^{-1}$, which are typical
 frequencies in infrared spectroscopy.\cite{basov} Similar effects were previously discussed in
 magnetotransport (in particular when a gap is opened in strong magnetic field).\cite{sharapov}

\section{Induced Vacuum Polarization Charge}

In the following  discussion, we assume $\omega=0$ ($q=|{\bf q}|$), and denote
$\Pi({\bf q}) = \Pi({\bf q},0)$.  First we extract
 the behavior in some  limiting cases. 
In the limit of large momenta we have
\begin{equation}
  \label{pol4}
  \Pi({\bf q}) = -q  \left ( \frac{1}{4} -  \frac{m^2}{q^2} \right), 
  \quad
  \frac{q}{m}  \gg 1 
  \,,
\end{equation}
whereas in the opposite limit
\begin{equation}
  \label{pol5}
  \Pi({\bf q}) = -\frac{q^2}{3\pi m},
  \quad
  \frac{q}{m} \ll 1
  \,.
\end{equation}
These limits determine the behavior of the polarization charge at small and large
 distances, respectively, which we proceed to investigate in more detail.
%
%

%

\begin{figure}[t]
  \centering
  \includegraphics*[width=0.47\textwidth]{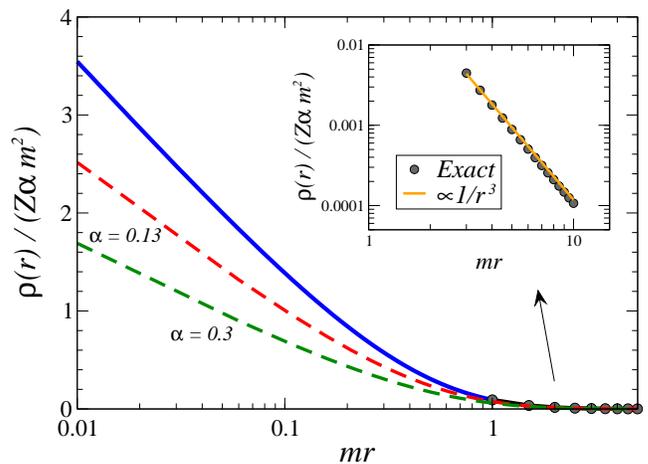}
  \caption{(Color online)
    Polarization charge from Eq.~\eqref{charge1} (solid line).
    The effect of additional electron-electron interactions (within RPA as
    discussed in the text), is also shown (dashed lines).
    Inset: magnification of the $1/r^{3}$ behavior at large
    distances.
  }
  \label{Fig2}
\end{figure}

The distribution of the polarization charge density in real space is given by
\begin{equation}
  \label{charge1}
  \rho(\br) = Z
  \int \frac{d^2\bq}{(2\pi)^2} V(\bq)\Pi(\bq) e^{i\bq.\br} 
  \,.
\end{equation} 
On the scale of the lattice spacing, $a$, away from the impurity there is a
contribution to the screening charge that reads
\begin{equation}
  \label{charge11}
  \rho(\br) = -Z\alpha \frac{\pi}{2} \ \delta(\br), 
  \quad r \simeq a  \rightarrow 0
  \,,
\end{equation}
which is valid in the continuum limit ($a\to 0$), and comes from the linear
contribution in \eqref{pol4}. In the massless situation ($m=0$) this localized
polarization charge is the final result. The finite mass introduces  new
behavior, namely additional polarization charge appears distributed in
space. The full behavior of $\rho(\br)$ is shown in Fig.~\ref{Fig2} (solid blue
line). One clearly identifies two distinct regimes whose asymptotic regions are
 determined by the Compton wavelength, the characteristic length scale of the
problem:
\begin{equation}
  \lambda_C = \frac{\hbar}{mv} \rightarrow \frac{1}{m} \ \ \ (\hbar=v=1)
\end{equation}
Using Eqs.~(\ref{pol4},\ref{pol5}), 
for distances much smaller than $\lambda_C$ (yet away from the Dirac-delta at
$r=0$) we find logarithmic decay,
\begin{equation}
  \label{charge2}
  \rho(\br) \sim  Z\alpha m^{2} 
  \; \ln{\left(\frac{1}{mr} \right)}, 
  \quad ma
    \ll mr \ll 1 
  \,,
\end{equation}
while at large distances a \emph{fast} power law emerges \cite{cheianov}
\begin{equation}
  \label{charge3}
  \rho(\br) \sim  Z\alpha m^{2} \ \frac{1}{(mr)^{3}} 
  \,, \quad mr \gg 1
  \,.
\end{equation}
These two regimes are evidenced in Fig.~\ref{Fig2} by means of the log-scale
in the main panel [cfr. Eq.~\eqref{charge2}] and by the fit shown in the inset
[cfr. Eq.~\eqref{charge3}].
The  numerical evaluation of Eq.~\eqref{charge1} shown in the figure
provides the crossover between the two asymptotic regimes. 
%

It is also significant to notice that the two contributions --- the
localized term \eqref{charge11}, and the spread tail ---  have different signs:
the lattice-scale contribution has a screening sign, while the long range tail
has a compensating,  \emph{anti-screening}, sign. This follows from 
the fact that, per Eq.~\eqref{pol5}, $\rho(q=0)=0$, meaning that the total
polarization charge $Q(\infty) = 0$, where $Q(R)$ is the vacuum charge
 accumulated within radius $R$:
\begin{equation}
  Q(R) = \int_{|\br|<R} \rho(\br) d\br  \,.
  \label{Qind}
\end{equation}
In fact, the rapid decay of $\rho(\br)$ beyond
$\lambda_C$ means that most of the additional (positive) charge, compensating
Eq.~\eqref{charge11}, is accumulated between the lattice scale ($r\simeq a$) and
$\lambda_C$, in such a way that $Q(R\gtrsim 1/m) \approx 0$.
This has peculiar consequences for screening: the impurity potential is
best screened at the shortest distances ($r\simeq a$), screening weakens
between  $a\lesssim r \lesssim \lambda_C$, and is essentially absent beyond
$\lambda_C$. Thus the impurity charge remains unscreened at large distances,
 as expected for an insulator.
%
\begin{figure}[t]
  \centering
  \includegraphics*[width=0.47\textwidth]{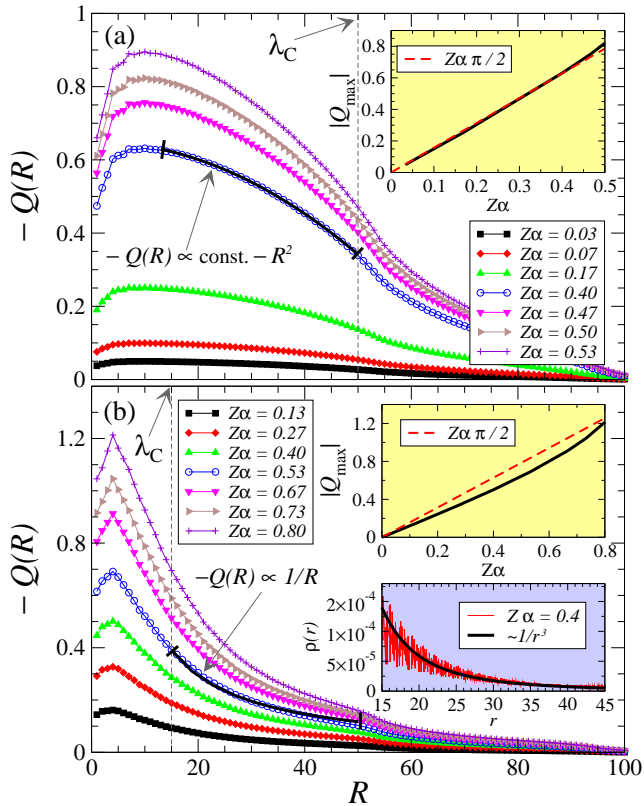}
  \caption{(Color online)
    Exact diagonalization plots of $Q(R)$, Eq.~\eqref{Qind}, 
    in the subcritical regime (we set $a=1$ in the plot). (a) Finite size gap only, with 
    $\lambda_C \simeq 50$. (b) Explicit mass $mv^2=0.1t$, leading to 
    $\lambda_C \simeq 15$. In both cases, the apparent inflections at 
    $R\simeq 55$ are due to boundary effects of the finite system.
    The top insets compare the maximum of $Q(R)$ ($Q_\text{max}$) obtained in
    the lattice with the value expected from Eq.~\eqref{charge11}. 
  } 
  \label{Fig3}
\end{figure}
%

%
\section{Induced Charge on a Finite Lattice}
To confirm the applicability of the above results to the real problem on a
lattice, and to dismiss possible regularization issues, we have investigated
via exact diagonalization the corresponding tight-binding model for graphene,
where the lattice appears naturally. The induced charge density can be
straightforwardly obtained with the aid of the exact wave-functions:
\begin{equation}
  \rho({\bf r}) = 
  - \sum_{E\le -m}\! \Psi_E^\dagger(\br)\Psi_E(\br)\!+\!
    \sum_{E\le -m}\! {\Psi_E^0}^\dagger(\br) \Psi_E^0(\br) 
  \,.
\end{equation}
Summation over the two spin components is also assumed (leading to an additional
 factor of 2). 
Here $\Psi_E^0$ are the wave-functions without external field ($Z=0$).
Rather than address the induced charge itself, it is more convenient
to consider  $Q(R)$, as defined in
Eq.~\eqref{Qind}. This quantity is already averaged over all directions
and is thus smoother and more appropriate for a direct numerical comparison.

We have studied two cases: (a) tight-binding model on a finite lattice with
$124\times 124$ sites, without explicit mass term, and (b) the same system with an
explicit mass of $mv^2=0.1t$, where $t$ is the hopping parameter. These two
cases allow us to address the two asymptotic regimes in Eqs.~\eqref{charge2}
and \eqref{charge3}. In  case (a), although $m$ is explicitly zero, there
is an effective gap due to the finite size of the system, $2 mv^2 =
\Delta\simeq 0.06 t$, and an effective $\lambda_C \simeq 50a$.\cite{vitor2}
Therefore, there is an appreciable region satisfying $a\lesssim r \lesssim
\lambda_C$. The calculated $Q(R)$ for this case is shown in Fig.~\ref{Fig3}(a).
Its behavior consists of a sharp increase for $R\sim a$ and a subsequent slow
decay up to $\lambda_C$. This decay follows the law $-Q(R) \propto (\mbox{const.} -R^{2})$,
 as one expects from
Eq.~\eqref{charge2}. Unfortunately, for our system $R=55a \approx \lambda_C$ is
the largest distance from the impurity that is free from boundary effects, and
one cannot comment on the crossover at larger distances. In order to address
that limit we look at case (b), for which $\lambda_C$ is much smaller 
($\lambda_C \approx 15a$); our results are shown in Fig.~\ref{Fig3}(b). 
In effect, we obtain $Q(R) \propto R^{-1}$
in the region $r\gtrsim \lambda_C$, in accordance with the result in
\eqref{charge3}. In the lower inset of the figure we show the $r^{-3}$ decay of
the actual induced charge on the lattice, for a particular value of the coupling.
The smallness of $\lambda_C$ in this case means that the intermediate,
logarithmic, regime is inaccessible. The analytical behavior thus stands in the
lattice problem, with qualitative and quantitative agreement, even for the case
when the gap is due to the finite lattice size.
\section{Role of Electron-Electron Interactions}
Electron-election interactions can influence the behavior described above.
Although questionable on account of the strict zero carrier density, one can perform
re-summation of  polarization loops within the random phase
approximation (RPA). For weak interactions RPA is expected
 to work quantitatively well, with deviations increasing as the
 interaction $\alpha$ increases.\cite{vertex}
 RPA amounts to the substitution 
  $
    \Pi(\bq) \rightarrow
    \Pi(\bq)
    /
    \bigl[
      {1-V(\bq)\Pi(\bq)}
    \bigr]
  $ 
in Eq.~\eqref{charge1}. The outcome of this procedure is given in
Fig.~\ref{Fig2} for different values of the interaction, being clear
that the result is a small decrease in the coefficient of the $\log$ in
Eq.~\eqref{charge2}. 

In addition, a qualitatively  important effect arises from self-energy corrections to the polarization. 
We evaluate the self-energy at  Hartree-Fock level, which was  first done for the massless case
in Ref.~\onlinecite{paco}.
In the massive case we obtain a velocity renormalization  of
%
\begin{equation}
  \label{hf}
  v \rightarrow v(q) = v\left \{1 +
  \frac{\alpha}{4}\ln{\left(\frac{\Lambda + 
  \sqrt{\Lambda^2 + m^2}}{q + \sqrt{q^2 + m^2}}\right)}\right \}
  \,, 
\end{equation}
where $\Lambda\sim 1/a$ is the ultraviolet cut-off, and the above expression is
 valid for $q,m \ll \Lambda$. The mass is also renormalized to a larger value $\tilde{m}$:
\begin{equation}
\label{mass}
m \rightarrow \tilde{m} = m\left\{1 + \frac{\alpha}{2}
\ln{\left(\frac{\Lambda}{m}\right)}\right\} , \ \ m \ll \Lambda.
\end{equation}
It is interesting to note that the logarithmic mass renormalization formula in graphene
\eqref{mass} is similar to the well-known expression for the electromagnetic mass (accounting for 
radiative corrections) in 3D relativistic QED.\cite{mass,iz}
 
From \eqref{hf}, for $\Lambda\gg q \gg m$, 
one has $v \rightarrow v(q)=v[1+\frac{\alpha}{4} \ln(\Lambda /q)]$, which 
leads to the ``running" of the coupling constant: $\alpha(q)=e^2/v(q)$.
Consequently, expanding $v(q)$ in powers of $\alpha$ leads, perturbatively, to
an additional piece in the polarization charge:
\begin{equation}
  \label{sub}
  \delta \rho(r) \sim \frac{Z\alpha^2}{16}\frac{1}{r^2}, \ \ a < r < 1/m
  \,.
\end{equation}
This is the perturbative limit of the effect, first  discussed in
Ref.~\onlinecite{sachdev}.

We conclude that electron-electron interactions affect somewhat the
above discussed behavior at the scale $\lambda_C$, but do not change the picture
qualitatively. This can be credited to the mentioned absence of screening, and
the fact that interactions do not generate an additional length-scale.
Furthermore, it seems that in current experiments $\alpha \ll 1$, making the
interaction corrections parametrically small.

\section{Conclusions}
We have found that in the presence of a finite mass gap the polarization charge, induced by a
Coulomb impurity in graphene, has a non-trivial behavior as a function of
distance. While at zero mass it is concentrated at the impurity site, at finite
mass it is distributed mostly up to $\lambda_C = 1/m$, with an additional power
law tail beyond that. Unlike the massless case, the total vacuum charge is zero
since the finite mass \emph{``pulls''} a compensating charge from
infinity to a distance  $\sim \lambda_C$, and the impurity charge remains unscreened
 beyond this scale.

In relativistic QED \cite{iz} the polarization charge at short distances has
anti-screening sign (enhances the potential), while the compensating 
charge is distributed up to the Compton wavelength of the electron. 
This behavior arises from the running of the charge $e^2(r)$ in QED. In
non-relativistic graphene (where the charge is not renormalized), the
situation is reversed as shown by the sign of Eqs.~\eqref{charge11} and
\eqref{charge2},\eqref{charge3}.

In the experiment of Ref.~\onlinecite{lanzara}, where the spectral gap is
$\Delta \approx 0.26 ~\text{eV}$, we have $\lambda_C \approx 30 a \approx 4
~\text{nm}$ and the behavior we predict in this work should be observable if an
external ion is present and generates the discussed charge re-distribution. In
practical terms, in order to  detect the density variation the chemical
potential might have to be at or above the value of the gap, $|\mu| \geq m$. Our
results will then be strictly valid only for $|\mu| \gtrsim m$, where the
screening length (determined by the occupation of the conduction band) remains
large and well separated from the scale $\lambda_C$. Although we strictly have
a hyperbolic band, when $|\mu|\gtrsim m$ we can resort to the behavior of a
parabolic band in 2D. Screening in this case is somewhat
peculiar \cite{ando} due to the finite density of states at the band
edge, $N(\mu=m) \propto m$. We expect that this could in fact facilitate 
detection of density variations via STM \cite{stolyarova,andrei} compared to the 
massless case in graphene. 

\begin{acknowledgments}
We are grateful to A. H. Castro Neto and O. P. Sushkov for
many  stimulating discussions.
V.\,M.\,P. was supported by FCT via SFRH/BPD/27182/2006.
\end{acknowledgments}


\bibliographystyle{asprev}

\end{document}